\documentclass[pdflatex,sn-mathphys-num]{sn-jnl}

\usepackage{graphicx}%
\usepackage{multirow}%
\usepackage{amsmath,amssymb,amsfonts}%
\usepackage{amsthm}%
\usepackage{mathrsfs}%
\usepackage[title]{appendix}%
\usepackage{xcolor}%
\usepackage{textcomp}%
\usepackage{manyfoot}%
\usepackage{booktabs}%
\usepackage{algorithm}%
\usepackage{algorithmicx}%
\usepackage{algpseudocode}%
\usepackage{listings}%
\usepackage{import}
\usepackage{array}
\usepackage{caption}
\usepackage{amssymb}
\usepackage{pifont}
\usepackage{subcaption}
\usepackage{tabularx}
\usepackage{setspace}
\usepackage{bm}
\usepackage{stfloats}
\usepackage{url}
\usepackage{verbatim}
\usepackage{makecell}
\usepackage{adjustbox}
\usepackage{academicons}
\usepackage{svg}
\usepackage{hyperref}
\newcommand{\cmark}{\textcolor{teal}{\ding{51}}}
\newcommand{\xmark}{\textcolor{purple}{\ding{55}}}

\theoremstyle{thmstyleone}%
\theoremstyle{thmstyletwo}%

\theoremstyle{thmstylethree}%

\begin{document}

\title[Article Title]{CRAFT: Laten\underline{c}y and Cost-Awa\underline{r}e Genetic-Based Framework for Node Pl\underline{a}cement in Edge-\underline{F}og Environmen\underline{t}s}


\author[1]{\fnm{Soheil} \sur{Mahdizadeh}}\email{soheil.mahdizadeh@sharif.edu}

\author[1]{\fnm{Amir Mahdi} \sur{Rasouli}}\email{mahdi.rasouli76@sharif.edu}

\author[1]{\fnm{Mohammad} \sur{Pourashory}}\email{mohamad.pourashory79@sharif.edu}

\author[1]{\fnm{Sadra} \sur{Galavani}}\email{sadra.galavani79@sharif.edu}

\author*[1]{\fnm{Mohsen} \sur{Ansari}}\email{ansari@sharif.edu}

\affil[1]{\orgdiv{Department of Computer Engineering}, \orgname{Sharif University of Technology}, \orgaddress{\city{Tehran}, \country{Iran}}}




\abstract{Reducing latency in the Internet of Things (IoT) is a critical concern. While cloud computing facilitates communication, it falls short of meeting real-time requirements reliably. Edge and fog computing have emerged as viable solutions by positioning computing nodes closer to end users, offering lower latency and increased processing power. An edge-fog framework comprises various components, including edge and fog nodes, whose strategic placement is crucial as it directly impacts latency and system cost. This paper presents an effective and tunable node placement strategy based on genetic algorithm to address the optimization problem of deploying edge and fog nodes. The main objective is to minimize latency and cost through optimal node placement. Simulation results demonstrate the proposed framework achieves up to 2.77\% latency and 31.15\% cost reduction.}




\maketitle

\section{Introduction}\label{sec1}
There are wide ranges of smart applications across different fields in the The Internet of Things (IoT) \cite{b1, b2}. High data rates in this technology present different challenges, including network congestion, missed deadlines, increased cost of implementation, etc. \cite{b3, b4}. To address these issues, various methods and technologies have been developed with edge and fog computing being among the most prominent \cite{b5,b6,b7,b8,b9}. \\
Edge computing emerged as a distributed computing paradigm designed to meet the latency-sensitive requirements of real-time applications by bringing computation and data storage closer to the network's edge. This proximity allows for efficient management of the vast amounts of data generated by IoT devices. The distributed nature enables some computation nodes to offload computational tasks from centralized data centers, thereby significantly reducing message exchange latency. Additionally, it helps balance network traffic, preventing traffic peaks and reducing transmission latency between edge nodes and end users, which is crucial for environments with real-time requirements \cite{b5, b6, b7}. \\
Fog computing is another paradigm that extends cloud computing capabilities by enhancing computation, communication, and storage closer to the network's edge. This approach supports latency-sensitive service requests with reduced energy consumption and lower traffic congestion compared to traditional cloud computing. Fog nodes can either process services using their available resources or delegate them to the cloud, thus enhancing resource efficiency and performance by reducing latency, conserving bandwidth, and lowering energy consumption \cite{b5,b6,b7,b8,b9}. \\
Minimizing end-to-end latency in IoT networks is the primary advantage of fog and edge architectures, especially in scenarios involving numerous IoT nodes and distributed fog nodes. The significance of these networks is evident in various applications, such as smart cities, healthcare, automated vehicles, and the industrial Internet of Things. Both edge and fog computing aim to bring storage, computing, and networking resources closer to the network’s edge, facilitating efficient data management and processing \cite{b8, b9}. \\
Despite the benefits, the practical implementation of these paradigms faces challenges that impact network efficiency. Implementing an IoT network using edge and fog computing paradigms involves several components, including Edge Nodes (ENs) and Fog Nodes (FNs). The placement and interconnection of these nodes significantly affect service latency \cite{b10, a1}. Improper distribution can increase implementation costs and cause load imbalances, where some nodes are overloaded while others remain underutilized or idle \cite{b11}. Thus, strategically placing ENs and FNs can significantly enhance application performance, reduce overall network latency, lower energy consumption, increase coverage, and minimize costs by reducing the number of required nodes \cite{b12}. \\
Motivated by these challenges, this paper presents a novel approach based on genetics algorithm to optimize node placement, thereby addressing issues related to latency and cost in IoT systems. Our approach consists of base stations which are interconnected forming a graph of nodes and links. These base stations act as candidates for ENs and FNs placement. Each user can connect to one EN and offload their computational needs. If the EN is overloaded, it offloads the users’ tasks to the nearest EN or FN based on network conditions and the computational capacity
of nodes. We also compared our algorithm to state of the art method, EVO \cite{b13}, to demonstrate its superior performance. \\
\subsection{Motivational Example}\label{AA}

\begin{figure*}[t]
\centerline{\includegraphics[width=\textwidth]{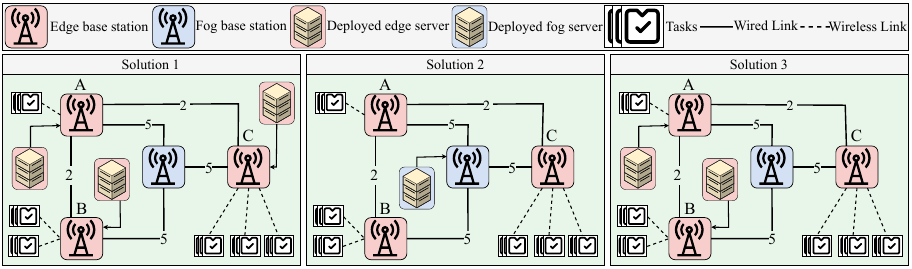}}
\caption{Motivational example of edge-fog environment node placement problem.}
\label{motivationalfig}
\end{figure*}

In this section, we present a motivational example to illustrate the potential cost-latency trade-offs in edge-fog environments.

\hyperref[motivationalfig]{Fig. 1} depicts three solutions for a scenario with three candidate edge base stations and one candidate fog base station. Each edge base station handles tasks from User Nodes (UNs) each one requiring one computation unit.

\textbf{Solution 1}: In the first solution, we deploy all three edge base stations. Here, each task is transmitted to the nearest server and executed locally, resulting in a low transmission latency. However, this configuration incurs a high total cost, equivalent to constructing three servers and six computation units.

\textbf{Solution 2}: In an alternative approach, we deploy a server with six computation units at the candidate fog base station. This configuration increases the average transmission latency to 5-time units due to the additional hop from the edge base stations to the fog tier. However, the total construction cost is significantly reduced to that of a single server.

\textbf{Solution 3}: A third option offers a balance between latency and cost. By placing servers at edge base stations A and B, the assignment of tasks to base stations resulted in a reduced average transmission latency of 2.3 time units. Specifically, the total latency was calculated as the sum of the transmission latencies, which is \(3 \times 2\) for base station C. However, to prevent the overloading of the edge server at base station A, one workload was redirected to base station B, incurring an additional latency of 2 time units. Consequently, the total latency increased to 8 time units, maintaining an average latency of 1.3 time units. The construction cost includes two servers and six
computation units, representing a moderate trade-off between
the two extremes.

It is important to note that edge base stations, such as Base Station C, must be equipped with additional access points to mitigate interference caused by a high number of connected users; otherwise, wireless transmission latency is likely to increase. These examples highlight the trade-offs between cost and latency in designing edge-fog environments, demonstrating how different configurations can impact performance and expenses.

\textbf{The main contributions of this paper are as follows:}
\begin{enumerate}
    \item Introducing an adaptive genetic algorithm for the placement of fog and edge nodes that outperforms other methods in the same scenarios and enables dynamic adjustments through the variable $V$. The parameter $V$ in this context is an adjustable argument within the genetic algorithm that influences the fitness evaluation. By altering $V$, the algorithm's focus can shift between prioritizing latency and cost.
    \item Providing a complete framework by considering different parameters such as latency, cost, wired and wireless communication, edge and fog nodes, etc.
    \item Offering flexible and scalable solutions for lower $V$ values (e.g., $V=10^4$) that minimizes placement cost but with higher latency, while for higher $V$ values (e.g., $V=10^6$), it prioritizes low latency at the expense of increased cost; and Finding a suitable $V$ value (e.g., $V=10^5$), balancing both latency and cost effectively.
\end{enumerate}

The remainder of the paper is structured as follows: \hyperref[relatedwork]{Section II} reviews related work and discusses proposed methods and their associated issues.  \hyperref[systemmodel]{Section III} presents our proposed system model and explains its components. \hyperref[proposedmethod]{Section IV} details the proposed genetic algorithm. Finally, \hyperref[evaluation]{Section V} evaluates the performance of the proposed method through various simulations and compares it with a baseline method to validate its effectiveness.

\section{Related Works}\label{sec2}
\label{relatedwork}
{
    \setlength{\aboverulesep}{0.1ex} 
    \setlength{\belowrulesep}{0.1ex} 
    \begin{sidewaystable}[htbp]
        \centering
        \caption{Comparison With Prior Works}
        \renewcommand{\arraystretch}{1.6}
        \begin{tabularx}{\textwidth}{>{\centering\arraybackslash}p{2cm}%
                                     >{\centering\arraybackslash}p{2cm}%
                                     >{\centering\arraybackslash}p{4cm}%
                                     >{\centering\arraybackslash}p{3cm}%
                                     >{\centering\arraybackslash}p{1.5cm}%
                                     >{\centering\arraybackslash}p{1.5cm}%
                                     >{\centering\arraybackslash}X}
            \toprule
            \textbf{References} & \textbf{Environment} & \textbf{Collaboration Between Layers} & \textbf{Heterogeneity} & \textbf{Latency} & \textbf{Cost} & \textbf{Wireless Communication} \\
            \toprule
            \cite{b14} & Fog & \cmark & \xmark & \cmark & \cmark & \cmark \\
            \cite{b15} & Fog & \xmark & \cmark & \xmark & \cmark & \xmark \\
            \cite{b16} & Fog & \cmark & \cmark & \cmark & \xmark & \cmark \\
            \cite{b17} & Fog & \xmark & \cmark & \cmark & \cmark & \cmark \\
            \cite{b18} & Fog & \cmark & \cmark & \cmark & \cmark & \cmark \\
            \cite{b19} & Edge & \cmark & \xmark & \cmark & \cmark & \cmark \\
            \cite{b20} & Edge & \xmark & \xmark & \cmark & \xmark & \cmark \\
            \cite{b21} & Edge & \cmark & \xmark & \cmark & \cmark & \cmark \\
            \cite{b22} & Edge & \cmark & \xmark & \cmark & \cmark & \cmark \\
            \cite{b23} & Edge & \xmark & \xmark & \cmark & \xmark & \xmark \\
            \cite{b24} & Edge & \xmark & \cmark & \cmark & \cmark & \cmark \\
            \cite{b25} & Edge & \xmark & \cmark & \cmark & \cmark & \cmark \\
            
            \cite{b26} & Edge-Fog & \cmark & \cmark & as QoS & as QoS & \cmark \\
            
            \cite{b27} & Edge-Fog & \xmark & \cmark & \cmark & \xmark & \cmark \\
        
            \textbf{CRAFT} & Edge-Fog & \cmark & \cmark & \cmark & \cmark & \cmark \\
            \bottomrule
            \vspace{0.1ex}
        \end{tabularx}
        \vspace{0.1ex}
        \begin{center}
            \footnotetext{
                (The parameters that have been taken into account are marked with a
                \cmark\ symbol, while those not considered are indicated by a
                \xmark\ symbol.)
            }
        \end{center}
        \label{table:compare}
    \end{sidewaystable}
}

The comparison of the proposed framework with various prior works in terms of different criteria is provided in \hyperref[table:compare]{Table I}. Most of the existing studies on the placement of nodes can be classified into three categories: fog, edge, and edge-fog environments. In terms of fog environments, there are several different works. For example, authors in \cite{b14}, proposed a multi-objective optimization approach using a memetic algorithm for efficient container placement in a fog computing environment. The goal was to provide a Pareto set of solutions for the container placement problem, ensuring efficient service deployment in an on-demand fog computing environment. Singh \textit{et al.} \cite{b15} proposed a model to address the fog node placement problem in a geographical area. The provided methodology uses a multi-objective genetic algorithm to optimize the placement of fog nodes by minimizing deployment cost and network latency. Faticanti \textit{et al.} \cite{b16} proposed a method for optimizing the placement of applications in a fog computing environment, specifically focusing on throughput awareness. The presented method involves partitioning applications into microservices and then placing these microservices onto fog nodes in a way that maximizes throughput while considering resource constraints such as CPU, memory, and storage. Ghalehtaki \textit{et al.} \cite{b17} proposed a Bee Colony-based algorithm for micro-cache placement in fog-based Content Delivery Networks (CDNs), where micro-caches are deployed on Set-top Boxes (STBs) near end users to reduce latency and improve end-user Quality of Service (QoS). The algorithm optimizes the placement of these micro-caches by balancing the latency and storage cost of each micro-cache, leveraging Network Function Virtualization (NFV) to manage caching functions. Manogaran \textit{et al.} \cite{b18} proposed an efficient resource allocation scheme for IoT-Fog-Cloud architectures that optimizes fog node placement to minimize service delays and resource exploitation. The architecture consists of heterogeneous fog nodes that serve as intermediaries between IoT devices and the cloud, providing localized processing to reduce latency and offloading tasks to the cloud when necessary. A profit function is used to determine the most efficient resource allocation, balancing the cost of fog node deployment with the need for high-quality, low-latency service.

Different aspects of edge environments are also investigated. Fan \textit{et al.} \cite{b19} introduced a cost-aware cloudlet placement in mobile edge computing algorithm, designed to optimize the placement of cloudlets in a network to balance deployment cost and end-to-end delay between users and their cloudlets. The proposed algorithm is a lagrangian heuristic designed to find a suboptimal solution to the cloudlet placement problem. Luo \textit{et al.} \cite{b20} addressed the edge node placement problem in mobile edge computing systems. The goal was to optimize the placement of edge nodes to balance the workload and minimize access delay between mobile users and edge nodes. The problem is framed as a multi-objective optimization problem and solved using a reinforcement learning approach. Authors in \cite{b21} proposed a method for placing edge nodes in a way that maximizes profit. The method utilizes particle swarm optimization with several enhancements to handle the specific requirements of edge node placement. The placement scheme aimed to balance the load across edge nodes, minimize latency, and maximize the overall profit. Ceselli \textit{et al.} \cite{b22} addressed the optimization of cloudlet placement in mobile access networks. The main focus is to design a network that supports virtual machine orchestration and user mobility while ensuring service level agreements are met. Authors in \cite{b23} investigated the problem of placing edge nodes within a specific geographic area to maximize the robustness of the edge node network. Robustness was defined as the ability of the network to maintain service when one or more nodes fail. The paper provided a comprehensive solution by combining an exact method suitable for small scales and an approximate method that performs well for larger scales, balancing computational efficiency with robustness optimization.
In \cite{b24}, Mazloomi \textit{et al.} proposed a reinforcement learning (RL)-based framework to optimize the placement of edge servers and allocation of workloads in Multiaccess Edge Computing (MEC). It formulates the joint problem as a combinatorial optimization problem to minimize both network latency and the number of deployed edge servers, reducing overall network costs. Kasi \textit{et al.} \cite{b25} proposed a multi-agent reinforcement learning (RL) approach to optimize the placement of mobile edge servers within Mobile Edge Computing (MEC) systems. The goal was to minimize latency and balance workloads across edge servers by leveraging a distributed, online learning model. The proposed solution aims to optimize edge server placement in real-time, addressing the heterogeneity of servers, and mobility considerations, while minimizing network delay and deployment costs.

In the edge-fog environments, Kochovski \textit{et al.} \cite{b26} proposed a method and architecture for automating the placement of database containers across edge, fog, and cloud infrastructures to ensure high Quality of Service (QoS). They addressed the cost and latency problem as part of their function for QoS. The method was based on markov decision processes and uses input data such as monitoring data from container runtime, expected workload, and user-related metrics to construct a probabilistic finite automaton that aids in both automated decision-making and placement success verification. Ma \textit{et al.} \cite{b27} addressed the problem of cloudlet placement in Wireless Metropolitan Area Networks (WMANs) aiming to minimize average access delay for mobile users offloading computational tasks. The authors propose two algorithms, a New Heuristic Algorithm (NHA), which improves efficiency by reducing redundant sorting processes during placement, and a Particle Swarm Optimization (PSO) algorithm designed for parallel execution to further enhance performance.

\section{System Model}\label{sec3}
\label{systemmodel}
As depicted in \hyperref[framework]{Fig. 2},  our framework consists of a number of base stations in a smart city that are randomly connected by fiber links forming a graph of nodes and links. These base stations are candidates for EN placement. There are also additional placement candidates for FNs. Each user node can connect to one EN using wireless communication and offload their computational needs. If the EN is overloaded, it offloads the users' tasks to the nearest EN or FN based on network conditions and the computational capacity of nodes. The details of the model are further explained in the subsections below:

\begin{figure*}[t]
\centerline{\includegraphics[width=\textwidth]{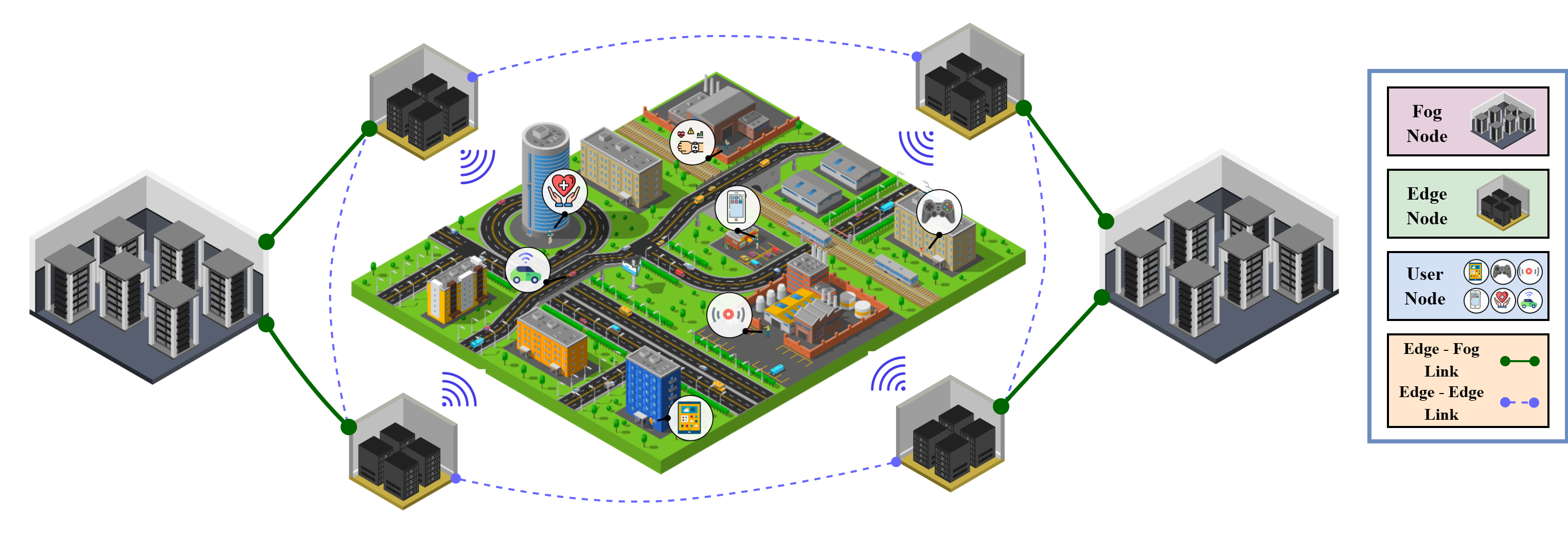}}
\caption{Our proposed framework for the edge-fog environment.}
\label{framework}
\end{figure*}

\subsection{Task Model}\label{AA}
In the task model for our system, each task (i.e., \( n_{i} \in N = \{ n_{1}, n_{2}, ..., n_{|N|}\} \)) is characterized by three primary properties (i.e., \( n_{i} = (d_{i}, freq_{i}, cycles_{i}) \)): data size, required computation frequency, and the number of CPU cycles necessary for execution as \( d_{i} \), \( freq_{i} \), and \( cycles_{i} \) respectively. \( d_{i} \) represents the tasks' data size needed to be processed. The computation frequency, indicating the processing speed needed to handle the task frequency is set as $freq_{i}$. This spectrum ensures that tasks are executed within an optimal time frame, balancing performance and resource utilization. The number of CPU cycles required for each task depends on its computational complexity and is considered to be $cycles_{i}$, directly influencing the execution time. By defining tasks with these properties, our model captures the diversity of real-world applications in an edge-fog computing environment, enabling precise allocation of resources and enhancing overall system performance and efficiency.

\subsection{Graph model}
\textit{\textbf{Network Link Model:}} We consider the randomly generated graph of nodes and links as a smart city map. As depicted in \hyperref[motivationalfig]{Fig. 1}, two kinds of nodes are present in the said graph which represent placement candidates for ENs or FNs which are denoted by $M$ and $F$ respectively. Every placement candidate can host one or more ENs/FNs to satisfy the needed maximum frequency in that area based on network conditions for latency and cost constraints. For each EN placement candidate, an EN named \(m_{j} \) with tasks to be executed directly from a user \(n_{i}\) or another EN (e.g., an overloaded EN) named \(m_{i}\) can be placed. For FN placement candidates, tasks only arrive from overloaded ENs to be executed. Ultimately, we have modeled our network links as a union of three explained subsets:
\begin{equation}
\begin{split}
E \: = \: & \{(n_{i},\: m_{j},\: \lambda_{ij})\: |\: n_{i} \in N,\: m_{j} \in M\} \: \cup \\             & \{(m_{i},\: m_{j},\: \lambda_{ij})\: |\: m_{i},\: m_{j} \in M\} \: \cup \\
         & \{(m_{i},\: f_{j},\: \lambda_{ij})\: |\: f_{i} \in F,\: m_{j} \in M\}
\end{split}
\label{eq:1}
\end{equation}
in which \( \lambda_{ij} \) is the bandwidth of the link between \( n_{i} \) and \( m_{j} \). Note that for each subset presented in set \( E \), the indices \(i\) and \(j\) differ with each other.

\textit{\textbf{Edge Node (EN):}} Servers for ENs can be deployed at various placement candidate locations, typically closer to the end-users to minimize latency and enhance real-time processing capabilities. Every edge node placement candidate location \( m_{i}: (sc_{i}, ac_{i}, x_{i}) \), \( x_{i} \in \{0, 1\} \) can host edge servers with every edge server having computational capacity modeled as frequency capacity set at fixed \( \alpha \) GHz. For adjusting maximum frequency capacity needed for a location, we consider \( sc_{i} \) as number of servers deployed at a location. This ensures that edge nodes can efficiently handle tasks requiring computational resources, providing immediate processing and response capabilities at the network's edge. The parameter \( ac_i \) represents the number of access points deployed at an edge node. This value is at least 1 and can be increased to reduce interference power within a spectrum. As a result, edge nodes with a higher number of UEs can maintain a high bit rate. We also considered \(x_{i}\) as a binary value to indicate whether at least one server is placed at that location.

\textit{\textbf{Fog Node (FN):}} servers for FNs, on the other hand, can be strategically placed within the network to handle more substantial computational tasks that require higher processing power. Each fog node placement candidate location (i.e., \( f_{i} : (sc_{i}, y_{i}) \), \( y_{i} \in \{0, 1\} \)) can have $sc_{i}$ servers with each server having its computational capacity set to \( \omega \) GHz. The parameter $y_{i}$ represents if at least one server is placed at $f_{i}$ or not.

\textit{\textbf{Deployment Strategy:}} The deployment of edge and fog nodes are conducted by selecting optimal placement candidates for each type (i.e., $m_{i}$ or $f_{i}$) with the number of servers and access points (i.e., $sc_{i}$ and $ac_{i}$) deployed.

\subsection{Latency Model}
In our latency model, the overall latency is a critical metric derived from two primary components: computation latency and communication latency. This comprehensive approach ensures an accurate assessment of the system's performance, essential for optimizing edge computing deployments.

\textit{\textbf{Computation Latency:}} The computation latency is determined by the processing capacity of the node in relation to the task's required frequency. Specifically, it is calculated by dividing the node's frequency capacity by the frequency needed to compute the task. This relationship ensures that higher-capacity nodes or tasks requiring lower frequencies result in reduced computation latency. The formula for computation latency \( T_{comp} \) is expressed as:
\begin{equation}
    T_{comp_{i}} = \frac{cycles_{i}}{freq_{i}} \label{eq:2}
\end{equation}
where $cycles_{i}$ represents the number of cycles needed to execute the task, and \( freq_{i} \) denotes the task's required frequency.

\textit{\textbf{Wireless Communication:}} There are three types of links in \( E \), as explained in Eq.~(\ref{eq:1}). The links between edge nodes and those between an edge node and a fog node are physical links with a constant bitrate. However, the links between users and edge nodes are wireless links, where the bitrate must be calculated based on the number of users attached and the number of deployed access points. To model the wireless bitrate, we first define interference and SINR. The mean interference is defined as:

\begin{equation}
   \bar{I_j} = 
        \begin{cases}
            (\frac{|\mathcal{N}_j|}{ac_j} - 1)\cdot P\cdot \bar{h} & \frac{|\mathcal{N}_j|}{ac_j} > 1\\
            0 & \frac{|\mathcal{N}_j|}{ac_j} \leq 1 
        \end{cases}
    \quad\quad    \forall{m_j} \in M
\end{equation}

where \( \mathcal{N}_j \) is the set of all users connected to edge node \( m_j \), \( P \) is the transmission power of UEs, and \( \bar{h} \) is the average interference power gain. Now, the SINR is expressed as:

\begin{equation}
   SINR_{ij} = \frac{P\cdot \bar{h}}{\sigma^2 + \bar{I}_j} \quad n_i\in N, m_j \in M
\end{equation}

in which $\sigma^2$ is the noise power. Furthermore, for each link between a UE and an edge node we have:

\begin{equation}
   \lambda_{ij} = W\cdot \log(1 + SINR_{ij})\quad(n_i, m_j, \lambda_{ij}) \in E
\end{equation}
where $W$ is the channel bandwidth.

\textit{\textbf{Transmission Latency:}} The transmission latency, reflects the time taken to transmit the task's data size over the network. It is obtained by dividing the task's data size by the link's bandwidth, ensuring that larger data sizes or lower bandwidths lead to higher transmission latencies. Usually, the computed tasks in the ENs/FNs are small in bit size so we only consider the transmission from users to ENs and ignore the transmission from ENs/FNs to users for simplicity. The formula for transmission latency \( T_{tr} \) is given by:
\begin{equation}
    T_{tr_{i}} = \sum_{hop_{ij} \in P_{k}} T_{hop_{ij}} \label{eq:3}
\end{equation}
\begin{equation}
  T_{hop_{ij}} = \frac{d_{k}}{\lambda_{ij}} \label{eq:4}
\end{equation}
where \( d_{k} \) represents the data size of the task\(_{k}\), \( \lambda_{ij} \) denotes the bandwidth of the communication link. \( P_{k} \) is a set consisting of links forming a path that task\(_{k}\) is transmitted through. The total transmission latency is the sum of every individual \textit{hop} a task takes (e.g. a user sends a task to an EN and it sends the task to an FN due to being overloaded and forming a total of 2 \textit{hops}).

\textit{\textbf{Total Latency:}} The total latency \( T_{total} \) of a task is the sum of the computation latency and the communication latency. This comprehensive latency measure captures the full delay experienced by the task from processing to transmission, providing a holistic view of the system's performance. The total latency is expressed as:
 \begin{equation}
  T_{total_{i}} = T_{comp_{i}} + T_{tr_{i}} \label{eq:5}   
 \end{equation}
The average latency of all users in the network is expressed as:
\begin{equation}
  \Bar{T} = \frac{\sum_{n_i \in N} T_{total_i}}{|N|} \label{eq:6}
\end{equation}
By calculating both computation and communication latency, our model offers a precise and reliable measure of task performance in an edge-fog computing environment.
\subsection{Cost Model}
Our cost model incorporates both the fixed costs associated with the initial construction of nodes and the dynamic costs related to the ongoing deployment of edge nodes at base stations.

\textit{\textbf{Fixed Costs:}} The fixed costs represent the initial investment required for constructing the nodes. This includes the expenses for hardware, setup, and initial configuration. These costs are constant and do not vary with the number of deployed servers.

\textit{\textbf{Dynamic Costs:}} The dynamic costs, on the other hand, are variable and depend on the number of servers deployed at base stations. These costs cover the expenses for installation, maintenance, and operational aspects of each additional server. The total cost of the system can be calculated as the sum of ENs and FNs costs:
\begin{equation}
    C_{total} = C_{m} + C_{f} \label{eq:7}
\end{equation}
With \( C_{m} \) representing the costs related to ENs and \( C_{f} \) denoting the total FNs costs:
\begin{equation}
   C_{m} = \sum_{m_{i} \in M} x_{i}(c_{fixed} + (sc_{i} + ac_{i}) \times c_{dynamic}) \label{eq:8} 
\end{equation}
\begin{equation}
   C_{f} = \sum_{f_{i} \in F} y_{i}(c_{fixed} + sc_{i} \times c_{dynamic}) \label{eq:9}
\end{equation}
in which \(c_{fixed}\) and \(c_{dynamic}\) are constant and their values are set according to \cite{b17}; \( sc_{i} \) is the number of servers and $ac_i$ is the number of access points deployed. The values \(x_{i}\) and \(y_{i}\) have the value of \(0\) or \(1\) indicating if a node is deployed as stated before. By incorporating both fixed and dynamic costs, our cost model provides a comprehensive view for the financial implications of deploying and expanding the infrastructure and ensures that the deployment strategy is both economically viable and scalable, promoting efficient resource allocation and sustainable growth in the implementation of edge-fog computing solutions.

\begin{sidewaystable}[htbp]
\centering
\captionsetup{justification=centering, labelsep=newline}
\caption{Main Notation Used Throughout The Paper}
\renewcommand{\arraystretch}{1.6}
\begin{tabular}{>{\centering\arraybackslash}m{1.5cm} >{\arraybackslash}m{7cm} | >{\centering\arraybackslash}m{1.5cm} >{\arraybackslash}m{7cm}}
    \hline
    \textbf{Symbol} & \textbf{Meaning} & \textbf{Symbol} & \textbf{Meaning} \\ \hline
    $N$ & Set of user nodes & $C_m$ & Total cost of edge node placement \\
    $M$ & Set of edge nodes & $C_f$ & Total cost of fog node placement \\
    $F$ & Set of fog nodes & $C_{total}$ & Total placement cost of the network \\
    $m_{i}$ & Placement location $i$ for edge node & $T_{hop_{ij}}$ & Transmission latency between nodes $i$ and $j$ \\
    $f_{i}$ & Placement location $i$ for fog node & $T_{tr}$ & Total transmission latency between nodes \\
    $freq_i$ & Required frequency to execute task of user node $i$ & $T_{comp_{i}}$ & Computation latency of task $i$ \\
    $d_i$ & Task data size of user node $i$ & $\gamma$ & Fixed frequency \\
    $cycles_i$ & Cycles needed to execute task user node $i$ & $\alpha$ & Fixed frequency set to edge/fog node \\
    $sc_i$ & Node capacity of edge or fog node $i$ & $\omega$ & Fixed frequency set to fog node \\
    $ac_i$ & Access point count of edge node $i$ & $\bar{h}$ & average power gain \\
    $x_i$ & Indicates if edge node $i$ is placed & $fitness(V)$ & Fitness function based on $V$ \\
    $y_i$ & Indicates if fog node $i$ is placed & $K$ & Size of the population \\
    $E$ & Set of all links in the network graph & $T$ & Number of generations \\
    $V$ & Tunable parameter & $\bar{I_j}$& Mean interference\\ \hline
\end{tabular}
\label{table:notation}
\end{sidewaystable}

\subsection{Problem Formulation}\label{SCM}
In this section, we formulate the problem of optimizing the cost and latency of deploying ENs and FNs in an edge-fog computing environment. Our goal is to jointly minimize the overall cost and latency, based on the tunable value $V$, through leveraging a genetic algorithm. The objective is to minimize a composite metric that combines both the cost and latency of the system. This metric is calculated through the fitness function, where the tunable parameter \(V \in [0, +\infty)\) determines the relative importance of cost and latency. If the users' required frequency exceeds the capacity of placed edge/fog nodes, the fitness function considers the solution invalid:
\begin{equation}
    \sum_{n_i \in N} freq_i \le \sum_{m_i \in M} x_i \times sc_i \times \alpha + \sum_{f_i \in F} y_i \times sc_i \times \omega
    \label{eq:10}
\end{equation}
Considering this constraint, the fitness function based on tunable parameter $V$, can be formulated as follows:
\begin{equation}
fitness(V) =
\begin{cases}
- (V \times \Bar{T} + C_{total}) & \small{\text{if (\ref{eq:10}) is } true} \; \\
-\infty & \small{o.w.}
\end{cases}
\label{eq:11}
\end{equation}
\makeatletter
\renewcommand{\ALG@name}{Algorithm}
\renewcommand{\algorithmicrequire}{\textbf{Input:}}
\renewcommand{\algorithmicensure}{\textbf{Output:}}
\makeatother

\begin{algorithm}
\caption{Proposed CRAFT Node Placement Based on Genetic Algorithm}\label{alg:1}
\begin{algorithmic}[1]
    \Require $E$, $K$, $T$, $V$
    \Ensure $Q_{\text{best}}$ 
    \State \textbf{Initialize:} Set $K$ individuals for the population randomly. \\
    Calculate the fitness value of each individual based on the fitness function (\ref{eq:11}). \\
    Calculate the diversity factor based on (\ref{eq:14}). \\
    Sort out the population based on each individual's fitness value.
    \For{$t = 1$ to $T$}
        \State Randomly choose two individuals as parents and perform the \textbf{adaptive crossover} operation based on diversity factor.
        \State Select the individuals from parents and offsprings for \textbf{adaptive mutation}  operation based on diversity factor.
        \State Calculate the fitness value of each new individual based on fitness function (\ref{eq:11}) and sort out the population.
        Calculate the diversity factor based on (\ref{eq:14}).
    \EndFor
    \State Select the best individual with highest fitness value as $Q_{\text{best}}$.
\end{algorithmic}
\end{algorithm}
Our main objective is to find a solution that maximizes the fitness function based on the tunable parameter $V$. Total cost (\(C_{total}\)) is a combination of fixed costs (i.e., initial node construction) and dynamic costs (i.e., deployment and maintenance of ENs, FNs and access points). The total latency (\(T_{total}\)) is the sum of computation latency and communication latency. Computation latency (\(T_{comp}\)) is determined by the tasks needed cycles divided by the task frequency, while communication latency (\(T_{tr}\)) is the task data size divided by the link bandwidth. The main notations used throughout this paper are gathered in \hyperref[table:notation]{Table II} for ease of reference.

\section{Proposed Method}\label{sec4}
\label{proposedmethod}
We solve the problem of deploying ENs and FNs by leveraging the formula (\ref{eq:11}). The initial population consists of chromosomes that involve a genome with edge and fog genes. Every gene is modeled as follows:
\begin{equation}
        Gene_{edge} \: = \: (m_{i},\: sc_{i}, \:ac_i, \: x_{i})
        \label{eq:12}
\end{equation} 
\begin{equation}
        Gene_{fog} \: = (f_{i},\: sc_{i},\: y_{i})
        \label{eq:13}
\end{equation}
In each gene related to ENs, $m_{i}$ is the name of the placement candidate for edge node in the original network graph, the parameter $sc_{i}$ represents the number of nodes placed at $m_{i}$, $ac_i$ is the number of access points delployed at $m_{i}$ and $x_{i}$ is a binary value (\( x \: \in \: \{0, \: 1\} \)) indicating if at least one edge node is placed (i.e., x = 1) or not (i.e., x = 0). For the fog node genes, it is the same as edge node genes with $f_{i}$ being placement candidate name for fog nodes, $sc_{i}$ being number of fog nodes placed, and $y_{i}$ being the node placement indicator, however there are no access points available at fog nodes because UEs are connected to edge nodes only. We can calculate the total frequency capacity of the edge or fog node using, \(sc_{i} \times \gamma\) while \( \gamma \: \in \: \{\alpha,\: \omega\} \). $\alpha$ and $\omega$ are fixed frequency values for one edge node and fog node respectively. So by altering the value of $sc_{i}$ in a tuple, we can ultimately set a needed frequency capacity in location $m_{i}$ or $f_{i}$. The CRAFT method, also utilizes the condition of the population. By calculating a diversity factor, CRAFT adapts its crossover and mutation to better search the solution space. The diversity factor is calculated as follows:
\begin{equation}
        DF = \frac{\max(fitness(V)) - \min(fitness(V))}{\max(\max(fitness(V)) - \min(fitness(V))}
        \label{eq:14}
\end{equation} 
which is essentially the normalized value of the fitness difference in the current generation. As shown in \hyperref[alg:1]{Algorithm 1}, the algorithm begins by initializing a population of chromosomes in line 1, with genes formulated at (\ref{eq:12}) and (\ref{eq:13}). In lines 2-4, the fitness of every chromosome is calculated using the formula (\ref{eq:11}) in the initial population. Additionally the diversity factor based on (\ref{eq:14}) is also calculated. If the constraint formulated in (\ref{eq:10}) is true, the total cost of that gene is calculated by (\ref{eq:11}) with the tunable value of $V$. If the value of $V$ is high, the parent selection process favors chromosomes with high \(C_{total}\) and low $\Bar{T}$; conversely, if $V$ is set to low, it favors chromosomes with low \(C_{total}\) and higher $\Bar{T}$. In lines 5-10, the algorithm proceeds with crossover and mutation operations. During crossover, parent chromosomes are combined to produce offspring by mixing genes from the selected parents to explore new potential solutions. Lower values of the diversity factor bias the crossover operator toward selecting individuals with higher fitness, thereby promoting exploitation. In contrast, higher values of the diversity factor encourage the selection of a broader range of individuals, thus enhancing exploration. Mutation introduces variability to the population by randomly altering the value of $x/y$, $sc_{i}$ or $ac_i$ in the edge or fog genes. To enhance adaptability, the mutation rate is increased when the diversity factor is low, encouraging exploration. However, when the population is diverse, the mutation rate is decreased to better exploit the existing solutions. After repeating this process for multiple generations, the chromosome with the highest value  (i.e., $Q_{best}$ in line 11) controlled by $V$ will be the best placement strategy for the city graph with location names (i.e., $m_{i}$ or $f_{i}$) and their respective frequency capacity and access points (i.e., $sc_{i}$ and $ac_{i}$).

\begin{table}[tb]
\centering
\captionsetup{justification=centering, labelsep=newline}
\caption{Simulation Configuration}
\renewcommand{\arraystretch}{1.3}
\begin{tabular}{
    >{\arraybackslash}m{4.5cm}
    >{\arraybackslash}m{2cm} |
    >{\arraybackslash}m{3.5cm}
    >{\arraybackslash}m{1.4cm}
}
\hline
\textbf{Device/Link Parameter} & \textbf{Value} & \textbf{Network/Genetic Parameter} & \textbf{Value} \\
\hline
Server fixed frequency ($\gamma$) & 0.5 GHz & Fog node count ($|F|$) & 5 \\
Fog node server count ($sc_i$) & [6, 8] & Edge node count ($|M|$) & 30 \\
Edge node server count ($sc_i$) & [4, 6] & User node count ($|N|$) & [70, 170] \\
Edge node access point count ($ac_i$) & [1, 5] & Population size & 1000 \\
Edge-edge bitrate ($\lambda_{ij}$) & [4.85, 6.85] Mbit/s & Generation count & 100 \\
Edge-fog bitrate ($\lambda_{ij}$) & [2.01, 4.01] Mbit/s & Mutation rate & [0.1, 0.3] \\
Average power gain ($\bar{h}$) & $10^{-5}$ & & \\
Noise power ($\sigma^2$) & -100 dbM & & \\
Required frequency of task ($freq_i$) & [50, 200] MHz & & \\
Required cycles of task ($cycles_i$) & [10, 60] cycle/bit & & \\
Data size ($d_i$) & [800, $4 \times 10^6$] bit & & \\
Fixed cost of placement ($c_{fixed}$) & 500 & & \\
Dynamic server/AP cost ($c_{dynamic}$) & 100 & & \\
Bandwidth ($W$) & 10 MHz & & \\
\hline
\end{tabular}
\label{table:parameters}
\end{table}

\section{Evaluation}\label{sec5}
\label{evaluation}
 In this section, we evaluated the proposed node placement framework to demonstrate its efficiency, scalability, and feasibility. For evaluation, we implemented a simulation environment based on the data of previous works \cite{b19, b28, b29, b30} and analyzed the simulation in various scenarios based on cost efficiency, network latency, and scalability of the proposed method. To best of our knowledge, the previous studies on edge-fog node placement have not provided a tunable algorithm that considers different environments and applications. As a result, in order to have a fair comparison, a random placement algorithm and the method provided in \cite{b13} called EVO was implemented as the baseline methods for comparison with the proposed framework.



\begin{figure}[htbp]
  \centering
  \begin{subfigure}[htbp]{0.7\textwidth}
  \centering
    \includegraphics[width=\textwidth]{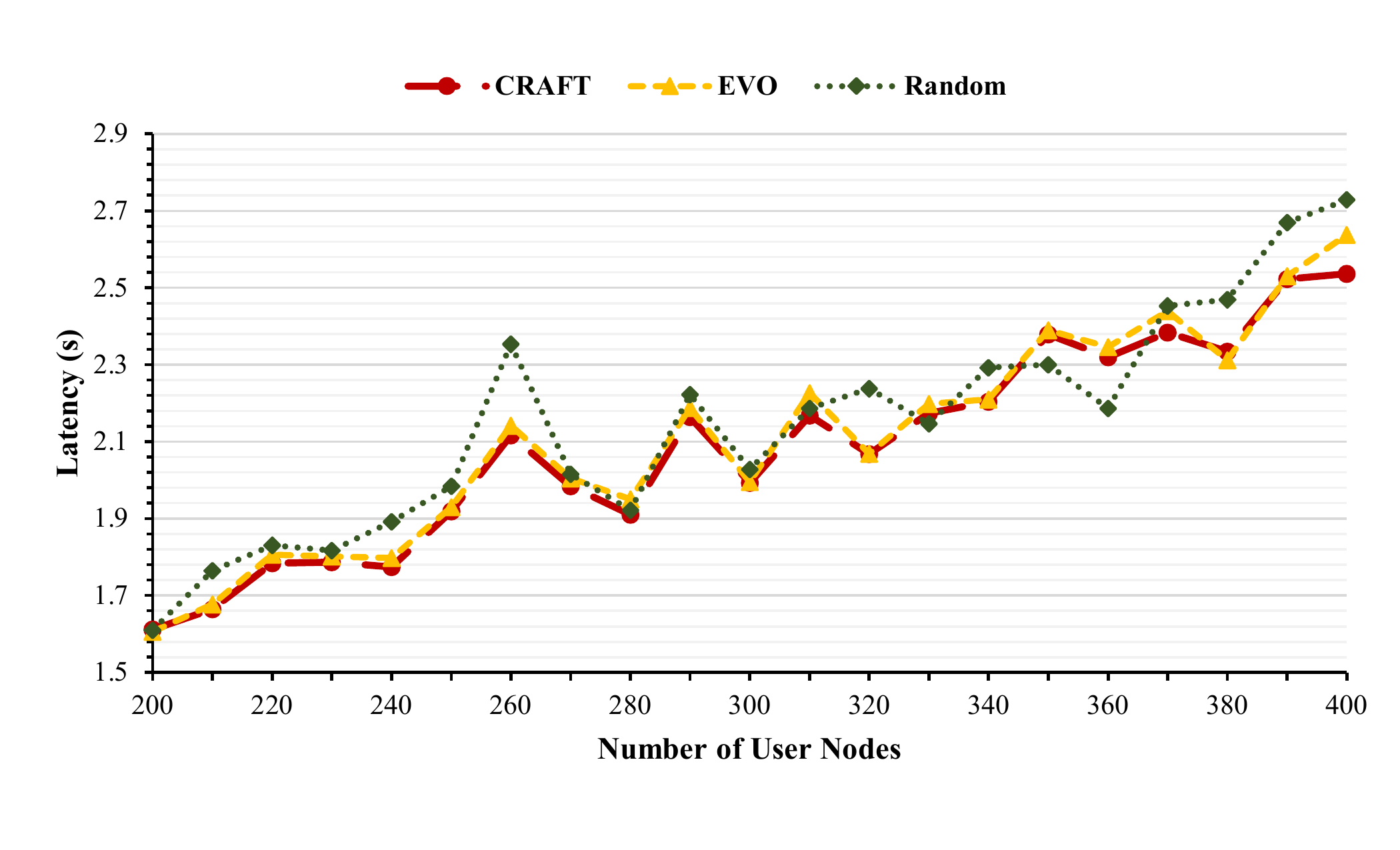}
    \caption{}
    \label{figplot1}
  \end{subfigure}
  \hfill
  \begin{subfigure}[htbp]{0.7\textwidth}
  \centering
    \includegraphics[width=\textwidth]{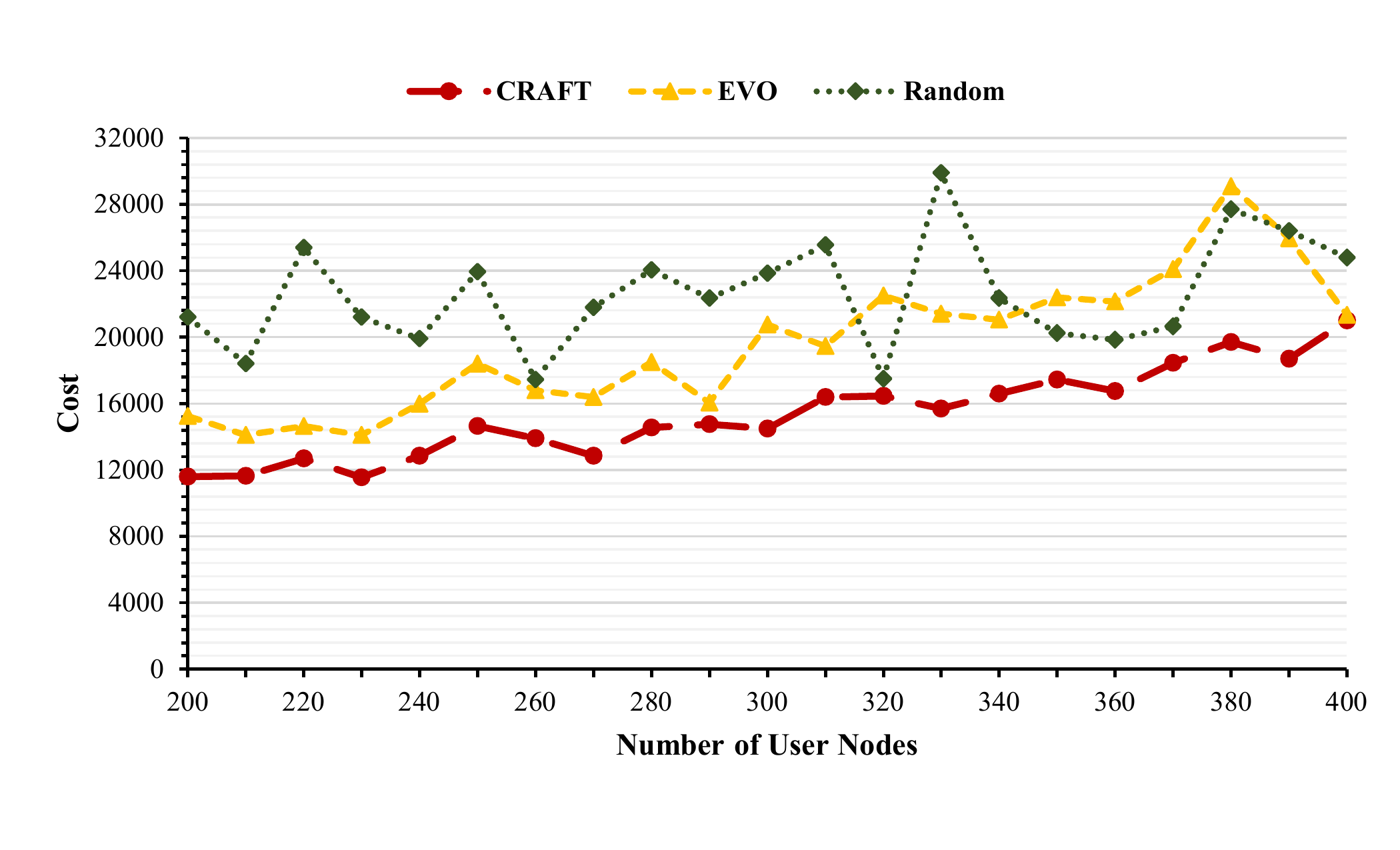}
    \caption{}
    \label{figplot2}
  \end{subfigure}
  \caption{Comparison of (a) Latency and (b) Cost for CRAFT, EVO, and Random approaches as the number of user nodes increases.}
  \label{fig:sidebyside}
\end{figure}

\begin{figure}[htbp]
\centerline{\includegraphics[width=0.7\textwidth]{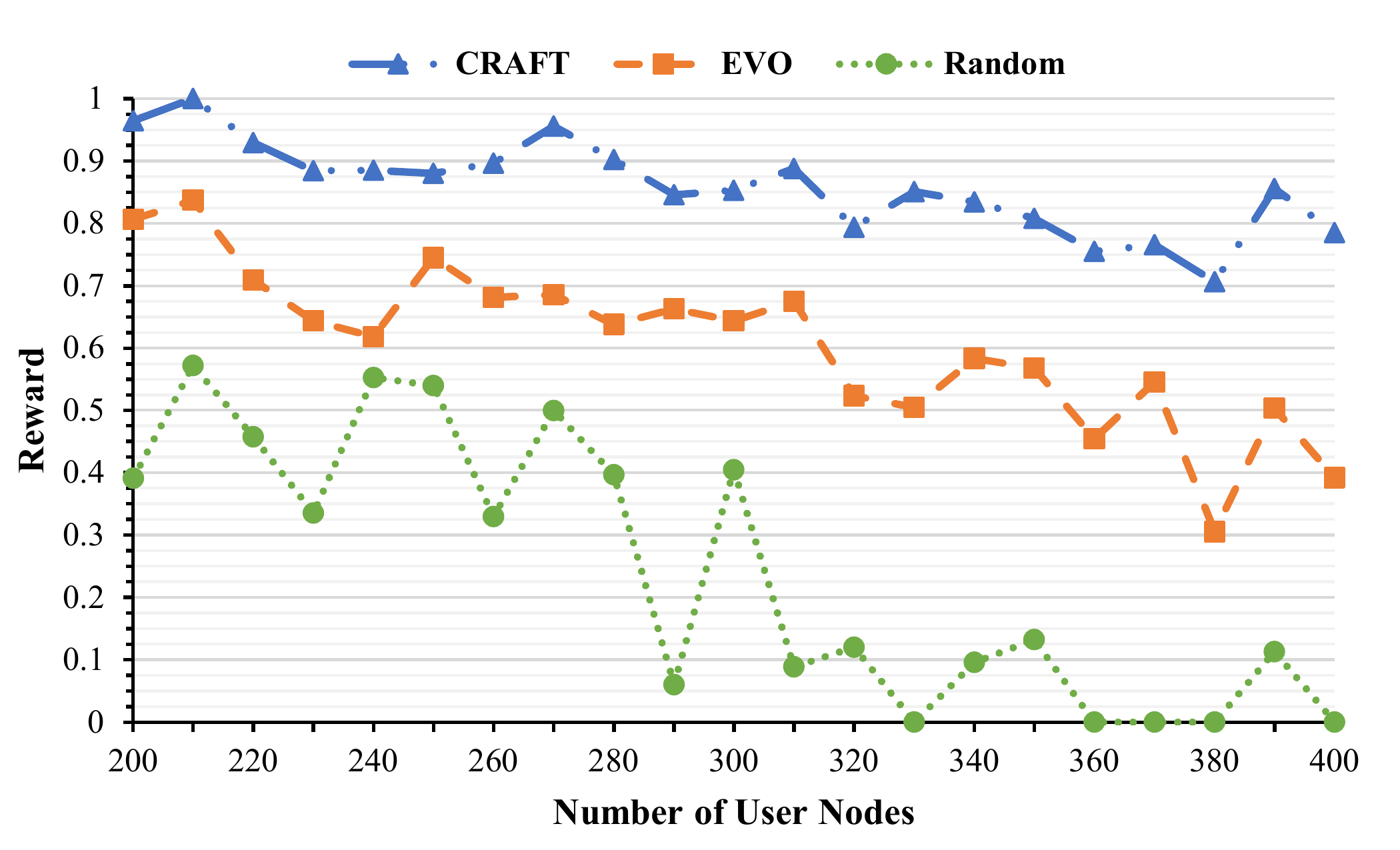}}
\caption{Comparison of reward performance versus the number of user nodes for CRAFT, EVO, and Random.}
\label{rewardnodes}
\end{figure}

\subsection{Simulation Environment}
We implemented a simulation environment for our proposed framework using Python3 programming language. Considering our tunable algorithm, we implemented this environment in an adjustable way. Our simulation configuration can be seen in \hyperref[table:parameters]{Table III}. 

\subsection{Evaluation and Results}
Our experiments consist of different scenarios. In the first scenario depicted in \hyperref[figplot1]{Figure 3 (a)}, we compare the latency efficiency of the network based on different numbers of user nodes and compare these results with the baseline. It can be observed that our CRAFT method offers lower levels of latency compared to the baseline and shows an average of 2.77\% improvement compared to random. In the second scenario shown in \hyperref[figplot2]{Figure 3 (b)}, the cost efficiency of our method is compared to other baseline methods, and our method can achieve lower cost for various numbers of user nodes. It can be seen that our CRAFT method offers 31.15\% and 20.64\% lower cost compared to random and EVO, respectively.

In \hyperref[rewardnodes]{Figure 4}, the comparative performance of CRAFT, EVO, and random across varying numbers of user nodes ranging from 200 to 400 is illustrated. Our CRAFT algorithm  consistently outperforms the other methods and maintains high reward values close to 0.9 even as the number of user nodes increases. The EVO method shows moderate performance. It starts with relatively high rewards, but its effectiveness declines gradually as the number of user nodes increases. The Random strategy  performs the worst. Its rewards fluctuate and show a downward trend as the number of user nodes increases.

\begin{figure}[htbp]
  \centering
  \begin{subfigure}[b]{0.7\textwidth}
    \includegraphics[width=\textwidth]{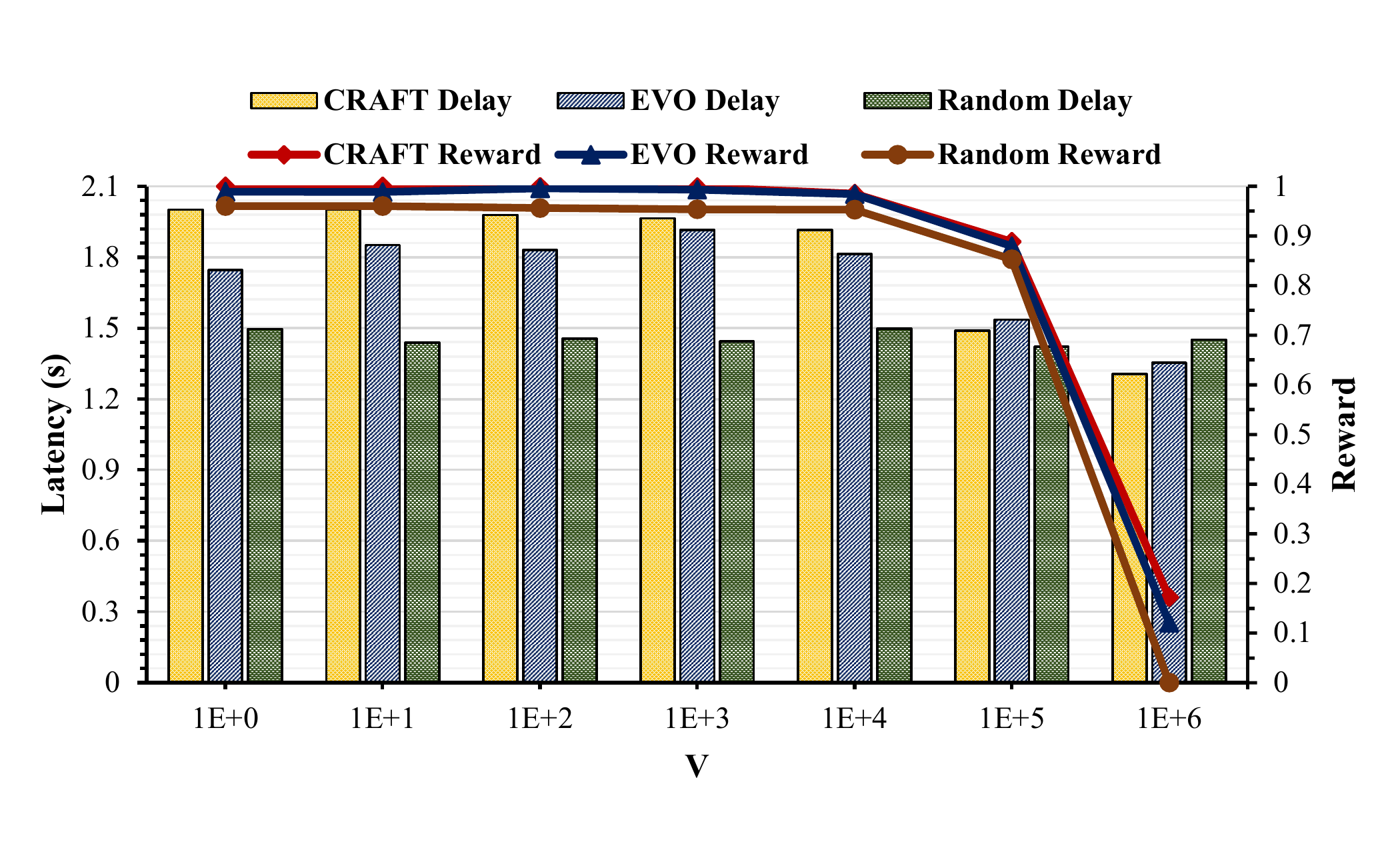}
    \caption{}
    \label{fig:plot1}
  \end{subfigure}
  \hfill
  \begin{subfigure}[b]{0.7\textwidth}
    \includegraphics[width=\textwidth]{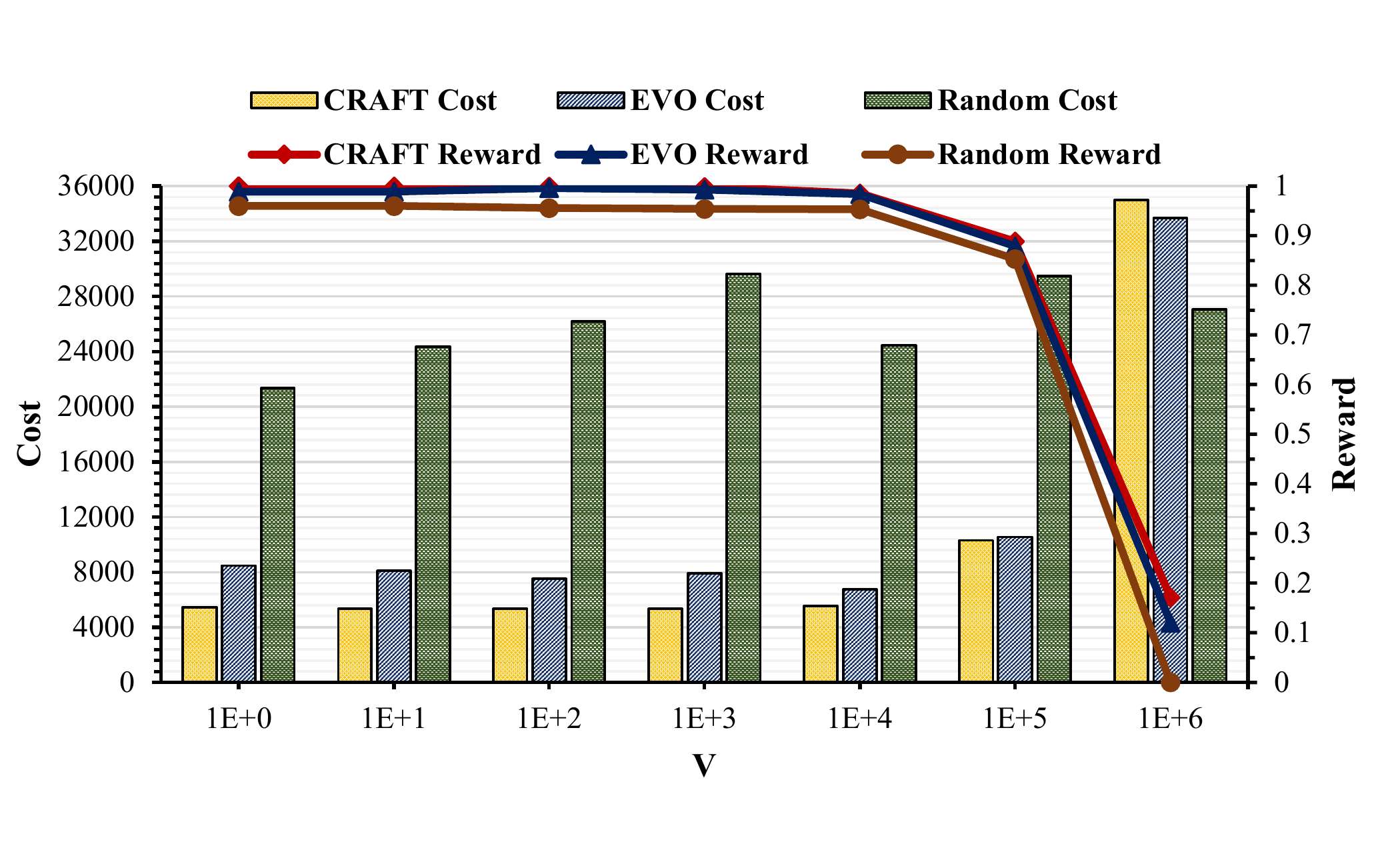}
    \caption{}
    \label{fig:plot2}
  \end{subfigure}
  \caption{Performance comparison of CRAFT, EVO, and Random algorithms under varying values of parameter $V$. (a) Average latency and reward versus $V$, (b) Total cost and reward versus $V$.}
  \label{fig:sidebyside}
\end{figure}

\hyperref[fig:plot1]{Figure 5 (a)} presents a comparison of latency and reward performance for the CRAFT, EVO, and random algorithms across varying values of the parameter $V$. The bar plots indicate the latency for each method. The line plots represent the 

corresponding reward values. At lower values of $V$, all three algorithms achieve high reward values close to 1, while the latency varies among them. At lower values of $V$, random achieves the lowest latency, followed by EVO, with CRAFT showing slightly higher latency. However, as the value of $V$ increases, the reward values for all algorithms begin to decline sharply, and random fails to yield any substantial reward.

This trade-off between latency and reward, controlled by the parameter $V$, reflects the balance each algorithm maintains between system responsiveness and performance optimization, and CRAFT maintains competitive reward levels while achieving acceptable latency, and shows its capacity to balance efficiency and performance more effectively across a wide range of $V$ values compared to the other approaches.
 
\hyperref[fig:plot2]{Figure 5 (b)} shows the relationship between cost and reward for the CRAFT, EVO, and Random algorithms across a range of values for the parameter $V$ from $10^0$ to $10^6$. Across lower values of $V$, all three algorithms achieve consistently high rewards, while demonstrating significant differences in cost. CRAFT consistently incurs the lowest cost, followed by EVO, with Random having substantially higher costs. This highlights CRAFT's superior cost-efficiency. As $V$ increases higher than $10^4$, the cost for CRAFT and EVO begins to rise, peaking at $V=10^6$, while random’s cost decreases but still fails to deliver meaningful reward.

\section{Conclusion}\label{sec6}
In this paper, we highlighted the importance of node placement to enhance overall network performance. The proposed genetic algorithm offers a novel solution by dynamically balancing cost and latency, catering to various application needs that allows for effective deployment in diverse scenarios. The provided algorithm is scalable, making it suitable for a wide range of IoT applications. Our algorithm's practical balance demonstrated better performance in simulations compared to baseline methods.

\bibliography{sn-bibliography}

\end{document}